\documentstyle[twocolumn,prl,aps]{revtex}
\begin{document}
\twocolumn[\hsize\textwidth\columnwidth\hsize\csname
@twocolumnfalse\endcsname

\title{Exact perturbative solution of the Kondo problem}
\author{P. Fendley and H. Saleur$^{*}$}
\address{Department of Physics, University of Southern
California,
Los Angeles CA 90089-0484}
\date{June 1995, cond-mat/9506104}
\maketitle
\begin{abstract}
We explicitly evaluate
the infinite series of integrals that appear in the
``Anderson-Yuval'' reformulation of the anisotropic Kondo
problem in terms of a one-dimensional Coulomb gas.
We do this by developing a general approach
relating the anisotropic Kondo problem of arbitrary spin
with the boundary sine-Gordon model, which describes
impurity tunneling in a Luttinger liquid and
in the fractional quantum Hall effect. The Kondo solution
then follows from the exact
perturbative solution of the latter model in terms of Jack
polynomials.

\end{abstract}
\pacs{PACS numbers:  ???}

]

\narrowtext
It was shown some 25 years ago that
the partition function of the anisotropic Kondo problem could be
expressed as a power series in the impurity coupling
\cite{YA,Schotte}.
Even though the coefficients
were given only as formal multi-dimensional integrals,
this led to progress in the understanding
of the Kondo problem.
For example, this enabled renormalization-group calculations to be
done and the phase structure determined \cite{AYH}.
In subsequent years, various additional aspects of the problem were
further understood by methods like the numerical renormalization
group \cite{Wilson}, expansions around the low-temperature fixed
point
\cite{Noz} and by the Bethe ansatz \cite{AFL,TW}. Despite all this
progress, the problem is still far from completely solved. For
example, the finite-temperature resistivity (the aspect which
motivated Kondo's analysis \cite{Kondo})
 has yet to be calculated
exactly. Moreover, the coefficients of the perturbative
 expansion still could not be evaluated
explicitly.

This latter aspect highlights a general puzzle of the
Bethe ansatz approach. While its results
are non-perturbative, only by crude numerical solution
does it provide much information on the
small-coupling
perturbative expansion (except when $T=0$).  The reason
is that for $T\ne 0$, the Bethe ansatz gives a solution in terms
of non-linear integral equations, and no technique is known
to solve them systematically around the non-interacting
fixed point.
In this paper, we describe a new formalism which avoids these
problems. We show how to
``do'' the Anderson-Yuval integrals,
by expressing them in terms of infinite
sums of ratios of gamma functions \cite{FLSjack}\ which can
be easily evaluated numerically.

We derive a simple
functional relation between the partition functions
of the Kondo
problem and another
well-known one-dimensional quantum problem, the boundary
sine-Gordon (BSG) model.
The derivation utilizes
some results of \cite{BLZ}, together with the realization that
the BSG model can be considered as a particular anisotropic Kondo
problem where the boundary spin is in a ``cyclic'' representation
of  the quantum group $SU(2)_q$.
The perturbative coefficients of both of these
partition functions can be
expressed as the partition function of a one-dimensional classical
``log-sine'' gas with
positive and negative charges which have logarithmic
interactions.
In the Kondo problem the charges must alternate in
sign in space, while in the BSG model
they may occur in any order.
With unrestricted ordering,
the partition function can be evaluated \cite{FLSjack}\
by utilizing various properties of
Jack polynomials \cite{Jack}.
Since the functional relation
gives the appropriate Kondo coefficients
simply in terms of the BSG ones, it therefore allows their
determination as well.

We first review the BSG and Kondo perturbative expansions.
We then derive the (non-perturbative) relation between
these partition functions, which gives one set of perturbative
coefficients in terms of the other.
Both sets have applications in a variety of physical problems.
The Kondo problem of course is realized in various
impurity compounds, and is also equivalent (in the anisotropic
case) to the dissipative quantum mechanics of a particle
in a double well \cite{Leggett}. The BSG model describes
tunneling in a Luttinger liquid (with application to
fractional quantum Hall edges) \cite{KF,FLSi}, while
in dissipative quantum mechanics it corresponds to an infinite
number of wells  \cite{Schmid}.

The boundary sine-Gordon model describes a free boson
$\phi(\sigma,t)$ on the half-line $\sigma\ge 0$ with
an interaction at the boundary $\sigma=0$.
We study the problem at non-zero temperature
$T$; in the path integral
this corresponds to a system where Euclidean time
(denoted by $\tau$) is a circle of periodicity $1/T$.
The bulk action is
\begin{equation}
S=-{1\over 4\pi g}\int_{0}^\infty d\sigma\int_0^{1/T}   d\tau\
[(\partial_\tau\phi)^2+(\partial_\sigma\phi)^2]
\label{bulk}
\end{equation}
while the boundary action is
\begin{equation}
S_B=
2\lambda\int_0^{1/T}
\ d\tau\ \cos\left[\phi(\sigma=0,\tau)\right].
\label{act}
\end{equation}
The coupling $\lambda$ is not
scale-invariant and interpolates between
 Neumann ($\lambda=0$) and Dirichlet ($\lambda\to\infty$)
boundary conditions on the boson.

Defining as usual the partition
function via the path integral
as $Z=\int[d\phi]\ e^{S+S_B}$ we introduce
${\cal Z}\equiv  {Z(\lambda)/Z(\lambda=0)}.$
We can rewrite this as a series by expanding out
$\exp(S_B)$ in powers of
$\lambda$. The coefficients
of this series are then integrals of correlators
in the free-boson theory.
Using the bosonic propagator on the edge of a half-cylinder with
Neumann
boundary conditions,
$$\langle\phi(0,\tau)\phi(0,\tau')\rangle=-2g
\ln\left|{\kappa \over\pi T}\sin {\pi T}
(\tau-\tau')\right|,$$
($\kappa$ is the inverse cutoff), one finds for example
$$ \langle e^{i\phi(0,\tau)}
e^{-i\phi(0,\tau')}\rangle =
\left|{\kappa\over\pi T}\sin{\pi T(\tau-\tau')}\right|^
{-{2g}}$$
Using Wick's theorem one finds the general correlators as
ratios of various ${\cal C}(\tau,\tau')\equiv
\left|2\sin{\pi T(\tau-\tau')}\right|^
{{2g}}$. The partition function
is the series
\begin{equation}
{\cal Z}=\sum_{n=0}^\infty x^{2n} Z_{2n}\qquad
x\equiv {\lambda\over T}
\left({2\pi T\over \kappa}\right)^{g}
\label{bsgsum}
\end{equation}
\begin{equation}
Z_{2n}={1 \over  (n!)^2}
\int_0^{1/T} \prod_{i} [d\tau_i d\tau_i']
 {\prod_{j<l} [{\cal C}(\tau_j,\tau_l)
 {\cal C}(\tau_j',\tau_l')]\over
\prod_{i,k}{\cal C}(\tau_i,\tau'_k)}
\label{bsgpert}
\end{equation}
where all indices run from $1$ to $2n$, and all integrals run
from $0$ to $1/T$ independently.
This expression requires regularization for $g\ge 1/2$,
but is well-defined otherwise.
The sum (\ref{bsgsum}) is the grand canonical partition
function for a classical two-dimensional Coulomb gas of
charged particles restricted to lie on a
(one-dimensional) circle
of circumference $1/T$. The charges are $\pm\sqrt{2g}$,
and the whole gas is electrically neutral;
$x$ is
the fugacity.

Explicit series expressions were found for the $Z_{2n}$
\cite{FLSjack}\ by expanding
the denominator of (\ref{bsgpert})
in terms of Jack polynomials \cite{Jack},
and using orthogonality relations of these polynomials.
Jack polynomials are indexed by a partition of an integer
${\bf m}\equiv (m_1,m_2, \dots...m_{\l({\bf m})}))$,
where $m_1\ge m_2 \dots \ge m_{\l({\bf m})}$.
Any ${\bf m}$ can be pictorially represented by a
Young tableau with $\l({\bf m})$ rows and
$m_j$ blocks in the $j^{\hbox{th}}$ row.
One finds \cite{FLSjack}\
\begin{eqnarray}
\label{part}
Z_{2n}=&& \left(
{
\Gamma({g} n+1)\over
n! \left[\Gamma(1+{g})\right]^n }
\right)^2 \times \\
&&\sum_{\bf m}
\prod_{i=1}^{\l({\bf m})}
\left[  {\Gamma({g}(n-i)+1)
\Gamma({g}(n+1-i)+m_i)
\over \Gamma({g}(n-i+1))
\Gamma({g}(n-i)+1+m_i) } \right]^2
\nonumber
\end{eqnarray}
where the sum is over all partitions ${\bf m}$ with
$\l({\bf m})\leq n$.
The sum converges only for $g<1/2$, which
is where the integrals in (\ref{bsgpert}) are
well-defined.

For $n=1$, the sum can be done explicitly (or the integral
can be done without Jack
functions) giving $ Z_2=
{\Gamma(1-2g)/\Gamma^2(1-g)} $.
Notice that the divergence at $g=1/2$
(the well-known ``free-fermion
point'') is just a simple pole, so we can analytically
continue around it to study behavior for $g>1/2$.
In fact, it turns out that at the free fermion point,
only the first coefficient in the
expansion of the free energy ($=-T\ln{\cal Z}$) diverges
making it possible to analytically continue all
the $Z_{2n}$ to $g>1/2$ \cite{us}.
The pole in $Z_2$ is a signal that there
are logarithmic terms in the perturbative expansion at $g=1/2$,
which can in
fact be computed exactly \cite{us}.
Thus this free-fermion point
(which corresponds to the Toulouse limit of the Kondo
problem) is pathological in some respects.

The perturbative expansion for the Kondo problem
found in \cite{YA,Schotte} is closely related.
The Kondo problem
is a three-dimensional non-relativistic
problem, with free electrons antiferromagnetically
coupled to a single fixed impurity.
By looking at $s$-waves only, we restrict
to the radial coordinate and this becomes a one-dimensional
quantum problem where
massless fermions $\psi_i(\sigma,t)$ ($i=1,2$ are the spin indices)
move on the half-line $\sigma\ge 0$ with a
quantum-mechanical spin $S_a$ at $\sigma=0$.
The interaction parameters are $I_z,I_+=I_-$; in the isotropic
case $I_z= I_+$. The
impurity action is
\begin{equation}
S_B=\sum_{i,j,a} I_a \int_0^{1/T} d\tau\ \psi^{\dagger}_i(0,\tau)
S_a \sigma^a_{ij}\psi_j(0,\tau)\label{pfouhh}
\end{equation}
where the $\sigma^a$ are the Pauli matrices.
By a well-known bosonization procedure \cite{Schotte},
this can be rewritten in
terms of a free boson $\phi(x,t)$
with bulk action (\ref{bulk}) and
$$S_B=I_+\int_0^{1/T} d\tau
\left
(S_+ e^{i\phi(0,\tau)} +h.c.\right),
$$
where we have absorbed $I_z$ in a redefinition of the coupling $g$.
The anisotropy is thus parametrized by $g$, with $g=1$ the isotropic
case and $g=1/2$ the Toulouse (free-fermion) limit.

In the perturbative expansion,
we get correlators of $\exp(\pm i\phi)$ like before.
The crucial difference arises from the $S_\pm$.
When the impurity has spin $1/2$, we have $S_+^2=S_-^2=0$,
and only terms of the form $S_+S_-S_+S_-\dots$ survive
in the perturbative expansion. In the one-dimensional gas, this
is the requirement that charges alternate in sign.
We define
the partition
function of the spin-$1/2$ Kondo problem  as
${\cal Z}_K=2Z_K(I_+)/Z_K(0)$, where the factor of $2$
ensures that the entropy is $\ln 2$ in the non-interacting limit
$I_+=0$.  Thus
$${\cal Z}_K(x_K)=2 +\sum_{n=1}^\infty
 (x_K)^{2n} Q_{2n}\qquad
x_K\equiv {I_+\over T}
\left({2\pi T\over \kappa}\right)^{g}.$$
The $Q_{2n}$ are the integrals
in (\ref{bsgpert}) times $2(n!)^2$,
but where the region of integration
is restricted to be
$0\le\tau_1\le\tau_1'\le\tau_2\le\dots\tau_n'\le 1/T$.
The periodicity of the integrand means that
$Z_2=Q_2$, but the others are different.

Our central result relates the
partition functions of the Kondo and BSG models. Defining
$q=\exp(i\pi g)$, we find that
\begin{equation}
{\cal Z}_K[(q-q^{-1})x] {\cal Z}(x)=
{\cal Z}(qx) + {\cal Z}(q^{-1}x).
\label{central}
\end{equation}
One simple check is that the equation is
consistent with the fact that ${\cal Z}(x)=1$ in the
isotropic case $q=-1$.
Plugging in the perturbative expansions gives the Kondo coefficients
$Q_{2n}$ in terms of the BSG coefficients $Z_{2n}$. For example
\begin{eqnarray}
4 \sin^2 (\pi g) Q_4 =&& -4\cos ^2(\pi g) Z_4 +(Z_2)^2
\label{forqiv}\\
16 \sin^6 (\pi g) Q_6 = &&\sin ^2(3\pi g) Z_6 +(Z_2)^3\sin^2 (\pi g)
\nonumber\\
&&  - \left(\sin^2 (2\pi g)+ \sin^2(\pi g)\right)Z_2Z_4\ .
\label{forqvi}
\end{eqnarray}
Using the series (\ref{part}), we can easily find
values for the $Z_{2n}$ and hence the $Q_{2n}$ for $g< 1/2$
by truncating the series
and summing it numerically.
For example, we have $Q_2=Z_2=\Gamma(1-2g)/(\Gamma(1-g))^2$, and

\begin{center}
\begin{tabular}{|c|c|c|c|} \hline
\multicolumn{4}{|c|}{\bf Values of Coefficents}\\ \hline
&{$g=2/5$} &{$g=1/3$} &{$g=1/4$}\\
\hline\hline
$Z_4$ &\ 1.910750624\ &\ 0.837804224\  &\ 0.4644013099\ \\ \hline
$Z_6$ &\ 1.088518710\ &\ 0.276783311\  &\ 0.0968299150\ \\ \hline
$Z_8$ &\ 0.439166887\ &\ 0.061847648\  &\ 0.0129159832\ \\ \hline
$Z_{10}$  &\ 0.135465650\  &\ 0.010210054\  &\ 0.0012208002\ \\
\hline\hline
$Q_4$ &\ 0.982706435\ &\ 0.432237451\  &\ 0.2322006549\ \\ \hline
$Q_6$ &\ 0.291860092\ &\ 0.074496009\  &\ 0.0242074788\ \\ \hline
$Q_8$ &\ 0.061852434\ &\ 0.008729195\  &\ 0.0016144979\ \\ \hline
$Q_{10}$  &\ 0.010067801\  &\ 0.000757798\  &\ 0.0000763000\ \\
\hline
\end{tabular}
\end{center}

\noindent As a check, some of the integrals
were explicitly evaluated using Monte Carlo methods;
the agreement is good. As discussed in \cite{FLSjack},
for $g$ rational there are relations
between the numbers beyond those in (\ref{central});
for example for $g=1/4$,
$Q_{2n}=Z_{2n}/2^{n-1}$. The knowledge
of the
first few coefficients $Q_{2n}$ provides a very good numerical
solution of the problem. The series can be extrapolated using Pade
approximants and
for instance the flow of the boundary entropy is reproduced fairly
accurately \cite{FLSjack}.

The proof  of (\ref{central})
requires first of all the introduction of the
``quantum-group'' algebra $SU(2)_q$, which is a
one-parameter deformation of the $SU(2)$ algebra
\cite{Sklyanin}. There are
 three generators $S_+,S_-$ and $S_z$, with  commutation
relations
\begin{eqnarray}
q^{S_z}S_+q^{-S_z}=q &&S_+ \qquad q^{-S_z}S_-q^{S_z}=qS_-
\nonumber\\
\left[S_+,S_-\right] &&=
{q^{2S_z}-q^{-2S_z}\over q-q^{-1}}.
\nonumber
\end{eqnarray}
Like $SU(2)$, the quantum-group algebra has representations
of any spin.
A careful analysis
shows that the general anisotropic Kondo problem
of arbitrary spin is integrable
only if the spins in (\ref{pfouhh}) obey the $SU(2)_q$ algebra.
For the
isotropic problem $q=-1$, the distinction is irrelevant
because $SU(2)_{-1}$ is identical to $SU(2)$.
The distinction is also irrelevant for any $q$
for spin $1/2$ or $1$,
because the the two-dimensional representation
of $SU(2)_q$ is given by the Pauli matrices as for $SU(2)$,
and the spin-1 representation
is also the same up to a rescaling of $S_+$ and $S_-$.
Other representations differ, however, so one must be careful
when discussing the anisotropic Kondo problem at higher spin.

Following \cite{BLZ}, let us now introduce the ``quantum monodromy
operator'' $L_j(x)$, which is defined as
\begin{eqnarray}
\label{fortj}
L_j(x)= {\cal T}\ &&\exp\left[I_+q^{-1}\int_0^{1/T}
\right. d\tau \\ \nonumber
&&\left( e^{2i\phi_L(0,\tau)}q^{S_z} S_+ +
\left. e^{-2i\phi_L(0,\tau)}q^{-S_z}S_-\right)\right]
\end{eqnarray}
Here  the ${\cal T}$ indicates time-ordering, and the $SU(2)_q$
generators are in the spin-$j$ representation. The field $\phi_L$ is
the
left-moving component of the field  $\phi$; with Neumann boundary
conditions we have $2\phi_L(0,\tau)=\phi(0,\tau)$.
The ``quantum transfer matrix''
$T_j \equiv \hbox{Tr }L_j$
is a fundamental part of the
Kondo problem: $\langle T_j(x)\rangle$
is identical to the spin-$j$ anisotropic Kondo perturbative
expansion.
In particular,
$${\cal Z}_K(x)=\langle T_{1/2}(x)\rangle .$$
This follows from expanding (\ref{fortj}) in powers of $x$
(the factors $q^{\pm S_z}$ cancel), and using the fact that
the vacuum  is an eigenstate of $T_j$.

As shown in \cite{BLZ}, the $L_j$ satisfy
the Yang-Baxter equation.
This results in a number of remarkable properties.
In particular, one finds that all of the
$T_j$ can be generated
from spin-$1/2$ via the relation \cite{BLZ}
\begin{equation}
T_j(q^{1/2}x)T_j(q^{-1/2}x)=1 + T_{j-{1\over 2}}(x)
T_{j+{1\over 2}}(x)
\label{fromblz}
\end{equation}
or via \cite{us}
\begin{equation}
T_{1\over 2}(q^{j+1/2}x)T_j(x)=
T_{j+{1\over 2}}(q^{1\over 2}x) + T_{j-{1\over 2}}(q^{-{1\over 2}}x)
\label{tobeproved}
\end{equation}
This gives
enough information to
relate the two partition functions ${\cal Z}$ and ${\cal Z}_K$
(and do much more).
When one
expands $T_j(x)$ for higher $j$
in powers of $x$, one obtains integrals with all sorts of charge
orderings, with weights depending on $q$ (because
$S_+$ and $S_-$ depend on $q$ for representations other than spin
$1/2$, and
because of the monodromy of the chiral fields).
For example, $S_{\pm}^3=0$ for spin 1, so
$++--+-$ appears at order $x^6$ in $T_1$, but $+++---$ does not.
Therefore, the unordered integrals $Z_{2n}$ in (\ref{bsgpert})
can be constructed by summing over appropriate combinations
of terms from the $T_j(x)$. Since the relations
(\ref{fromblz}) or (\ref{tobeproved}) give all of the higher
$T_j(x)$ in terms of $T_{1/2}$, this means that any $Z_{2n}$
can in principle be expressed in terms of the $Q_{2n}$.
The perturbative results
(\ref{forqiv}) and (\ref{forqvi})
follow directly from (\ref{fromblz}).

The quickest way to derive (\ref{central})
is to introduce cyclic representations
of $SU(2)_q$. These occur when $q$ is any root of
unity, $q^t=\pm 1$, and have no analog in ordinary
$SU(2)$. They
are indexed by an arbitrary
complex parameter $\delta$ and
have $t$ states $|m\rangle$, $m=0,\ldots,t-1$
with generators acting as
$$S_{\pm}|m\rangle=
{q^{\delta\mp m}-q^{-\delta\pm m}\over q-q^{-1}}
|m\pm 1\rangle\qquad
S_z|m\rangle=m|m\rangle
$$
where states $|m\rangle$ and $|m+t\rangle$ are identified. Notice
that
in such a representation one can consider arbitrary powers of the
generators. Referring to cyclic representations as
spin-$\delta$, one finds that
all previous results for
spin-$j$ apply.  In particular, relation (\ref{tobeproved})
holds with $j\rightarrow\delta$. In
the formal limit where $\delta=-i\Delta$ with $\Delta>>1$ we have
$$
S_{\pm}\approx {e^{\pi g\Delta}q^{\mp m}\over q-q^{-1}}.
$$
In this limit the commutator $[S_+,S_-]$ becomes negligible, so
the traces of all monomials with $n$ generators $S_+$ and $n$
generators $S_-$ become identical, e.g.\
$$\hbox{Tr} (S_+S_-)^n=t\left(
{q^{1/2}e^{\pi g\Delta}\over q-q^{-1}}\right)^{2n}.$$
(One might have been tempted to study
$\lim_{j\to\infty} T_j$
where arbitrary powers of the generators also occur,
but this is not ${\cal Z}$ because
the traces do not behave in the
right way.)  Hence one way of expressing ${\cal Z}$ is
\begin{equation}
{\cal Z}(x)={1\over t}
\left<T_{-i\Delta}\left[ (q-q^{-1})q^{-1/2}e^{-\pi
g\Delta}x\right]\right>_{\Delta>>1}.
\label{zcyclic}
\end{equation}
We substitute $j\rightarrow\delta=-i\Delta$
and $x\rightarrow (q-q^{-1})q^{-\delta-1/2}x$
into  (\ref{tobeproved}). Letting $\Delta>>1$
proves (\ref{central}) for $q$ any root of unity (rational $g$).
The result follows
for any $|q|=1$ (real $g$) by continuity.
The renormalization of the Kondo coupling in
(\ref{central}) can be checked through the  identity $Z_2=Q_2$.
This renormalization ensures that (\ref{central}) makes sense
in the isotropic limit $q\to -1$, where the
Kondo partition function is a non-trivial object but ${\cal
Z}(x)=1$.

Having obtained the fundamental relation (\ref{central}),
higher-spin
Kondo partition functions follow from fusion, using the
relations (\ref{fromblz}) or (\ref{tobeproved}).
For instance one has
\begin{eqnarray}
\nonumber \langle
T_1\left[(q-q^{-1})x\right]\rangle
{\cal Z}(q^{1/2}x){\cal Z}(q^{-1/2}x)=\\
{\cal Z}(q^{3/2}x){\cal Z}(q^{1/2}x)
+h.c.+
{\cal Z}(q^{3/2}x){\cal Z}(q^{-3/2}x).
\end{eqnarray}
Other such relations give
integrals of the form (\ref{bsgpert}) with any charge ordering.
Slightly more complicated ordered integrals
arise in many places, and we hope our results
can be extended. One possibility is
in the Keldysh calculation of the
non-zero-frequency noise in the BSG model \cite{CFW}.
Another is the quantity  $P(t,T)$
in the double-well problem of dissipative quantum mechanics, which
is the probability that a particle at temperature $T$
is in one well at time $t$ given that
it is localized in that well
at $t=0$ \cite{Leggett}.
We have found the curious result that $P(iT,T)=
T_1/2-1/2$,
but its significance is not clear.

We have therefore succeeded in computing explicitly the
Coulomb-gas integrals for the anisotropic Kondo problem.
Our approach suggests a very deep structure,
where the integral equations
of the Bethe ansatz can be solved
in terms of Jack symmetric functions and quantum-group
combinatorics. We hope
that further studies will uncover more features of these relations.

\vskip .2in
\noindent $^*$ Packard Fellow.
\vskip .2in
This work grew out of a conversation with Al.B.\ Zamolodchikov, for
which we are grateful. We thank F.\ Lesage for doing the Monte Carlo
integrals, and for helpful discussions.
This work was supported by the Packard Foundation, the
National Young Investigator program (NSF-PHY-9357207) and
the DOE (DE-FG03-84ER40168).

\end{document}